\documentclass[%
 aip,
 amsmath,amssymb,
 reprint,%
]{revtex4-2}

\usepackage{graphicx}
\usepackage{dcolumn}
\usepackage{bm}

\usepackage[utf8]{inputenc}
\usepackage[T1]{fontenc}
\usepackage{mathptmx}
\usepackage{etoolbox}

\makeatletter
\def\@email#1#2{%
 \endgroup
 \patchcmd{\titleblock@produce}
  {\frontmatter@RRAPformat}
  {\frontmatter@RRAPformat{\produce@RRAP{*#1\href{mailto:#2}{#2}}}\frontmatter@RRAPformat}
  {}{}
}%
\makeatother
\begin{document}

\preprint{AIP/123-QED}

\title[Nonlinear dynamics]{Characterization of Nonlinear Dynamics in Semiconductors in Frequency Domain using Modulated Photoexcitation}
\author{Attia Awan}

\author{Rong Tang}%

\author{Zhou Kang}


\author{Khadga Jung Karki}
\affiliation{%
Physics Department, Guangdong Technion-Israel Institute of Technology, Shantou 515063, China
}%
\altaffiliation{Guangdong Provincial Key Laboratory of Materials and Technologies for Energy Conversion,
Guangdong Technion-Israel Institute of Technology, Shantou 515063, China}
\email{khadga.karki@gtiit.edu.cn}

\date{\today}

\begin{abstract}
Carrier dynamics in semiconductors is inherently complex owing to the coexistence of different excited species, such as free carriers and excitons, and their interactions among themselves and with traps and phonons. It is usual to use time-resolved responses to identify the processes that contribute to the dynamics. However, the responses often are non-exponential, leading to ambiguity in the interpretation. Here, we propose a frequency domain method to characterize nonlinear dynamics in semiconductors using CdSe as the test system. We show that by analyzing the frequency components in the total photoluminescence induced by a pair of phase-modulated beams, the parameters, such as rates of different types of recombination of free carriers and excitons,  can be evaluated. The method can be used as a simple diagnostic tool to characterize ultrafast processes that are relevant to the functionality of semiconductor devices.    
\end{abstract}

\maketitle

 Analyzing harmonic responses to sinusoidal perturbations is a standard method to characterize circuit nonlinearities.\cite{WAMBACQ2010} While linear systems preserve the input frequency, nonlinearities distort the waveform, generating harmonics at integer multiples of the fundamental frequency. Standard electrical techniques, however, cannot probe timescales below tens of picoseconds, limiting their use to slower processes like thermal or trapping memory effects.\cite{YA2016}

In optoelectronic systems, dynamics span a wide temporal hierarchy. Light absorption triggers sub-picosecond physical processes, including carrier thermalization, carrier-carrier/carrier-phonon scattering, Auger, and radiative or trap-assisted recombination.\cite{SHAH1993,DESHLER2017,LOH2014,SCHENK1992,KARKI2012} Conversely, macroscopic device responses like photocurrent and photoluminescence (PL) occur on nanosecond to microsecond timescales. This temporal disparity presents a core modeling challenge: slower downstream dynamics (transport, drift, diffusion, and capacitance charging) act as a low-pass filter that averages out ultrafast optical phenomena. Preserving sub-picosecond signatures in these slow signals is essential for optimizing high-speed photodetection,\cite{SOLJAVIC2009} all-optical switching,\cite{GONG2021,CHITRA2018} and nonlinear optical signal processing.\cite{CLADER2014,OMER2014}

In a pivotal early demonstration, Bergqvist \textit{et al.} showed that fast dynamics leave measurable imprints on slow responses by using two intensity-modulated laser beams to isolate bimolecular recombination via photocurrent intermodulation products.\cite{INGANAS2016} Similarly, fast nonlinearities can manifest as unwanted artifacts during two-dimensional optical spectroscopy of perovskites.\cite{SILVA2017, SILVA2023, MORAN2023, FRESCH2023, KARKI2019} These spurious signals stem from incoherent nonlinear processes that distort coherent features, yet analyzing them offers a unique pathway to probe carrier transport, trapping, and recombination.\cite{MORAN2023,KARKI2019B}

\begin{figure}
\includegraphics[width=1\linewidth]{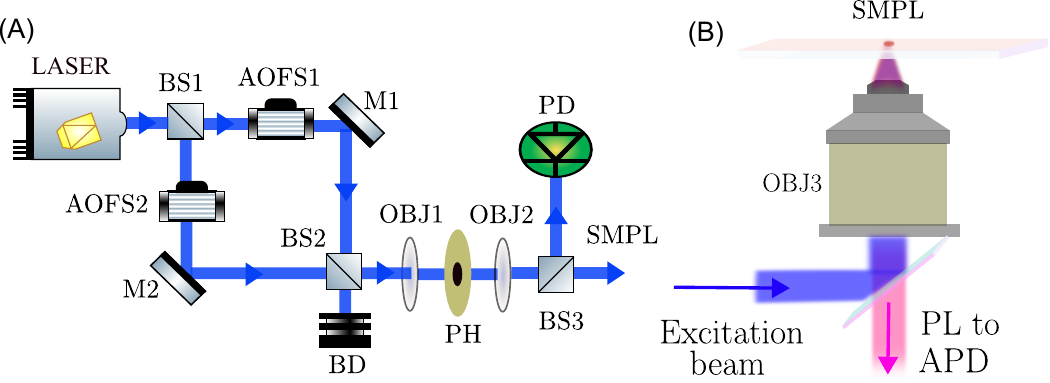}
\caption{\label{fig:1} (A) Schematic of the setup for single frequency modulation of the laser intensity. The output of a CW laser at 450 nm is directed to a Mach-Zehnder interferometer via a beamsplitter (BS). Acousto-optic frequency shifters (AOFs) in each arm of the interferometer shift the frequency of the beam by their respective driving frequency. The two beams are recombined collinearly by a second beamsplitter. One of the outputs of the interferometer is used as the modulated light source. Mode filtering by tightly focusing the beam with a microscope objective (OBJ1) to a 10 $\mu$m aperture pin-hole and recollimating by a second objective (OBJ2) is applied to maximize the depth of modulation. A beamsplitter further splits the beam into two, one of which is directed to a photodiode for reference, and the other is directed to an inverted microscope to excite the sample (SMPL). (B) The microscope is equipped with a long-pass dichroic mirror with the edge at 550 nm. A 10X objective focuses the beam onto the sample, and the resulting PL in the epi-direction is collected by the same objective and filtered by the dichroic mirror. Additional long-pass filters are used to suppress the scattered light. The PL is detected by an avalanche photodiode (APD). }
\end{figure}

As direct laser modulation using electro-optic or mechanical means inherently introduces strong excitation overtones,\cite{INGANAS2016} only intermodulation frequencies provide clean data, while harmonic signals remain heavily contaminated by source overtones that cannot be easily eliminated. This limitation can be bypassed by interfering two phase-modulated beams. Proper phase modulation produces single-tone intensity modulation free of overtone content\cite{BRUDER2015,KARKI2016} as has been demonstrated using femtosecond pulses to track coherent dynamics.\cite{KARKI2019B} Extending this method to continuous-wave (CW) excitation enables overtone-free isolation of intrinsic material nonlinearities, providing clean characterization of recombination dynamics. Here, we demonstrate this principle via PL tracking on bulk CdSe. To achieve single-tone modulation, a laser output is split into two equal, phase-coherent paths, and a slight frequency difference is introduced via phase modulation.\cite{WARREN_2002} A schematic of the optical setup is shown in Figure~\ref{fig:1}.

 In the optical setup, a CW laser operating at a wavelength of 450 nm serves as the light source. The laser output is split into two paths using a Mach–Zehnder interferometer configuration. Each arm of the interferometer is equipped with an acousto-optic frequency shifter (AOFS), which imposes phase modulation on the respective beam at frequencies of $f_1 = 54.995$ MHz and $f_2 = 55$ MHz. Consequently, the optical frequencies of the two beams are shifted to $\omega_1 = \omega + 2\pi f_1$ and $\omega_2 = \omega + 2 \pi f_2$.
 When the beams are recombined, the resulting interference produces an intensity modulation at the angular frequency $\phi = 2\pi (f_1-f_2)$, described by $I(t) = I_0 \left(1 + \cos(\phi t)\right)$.

\begin{figure}
\includegraphics[width=1\linewidth]{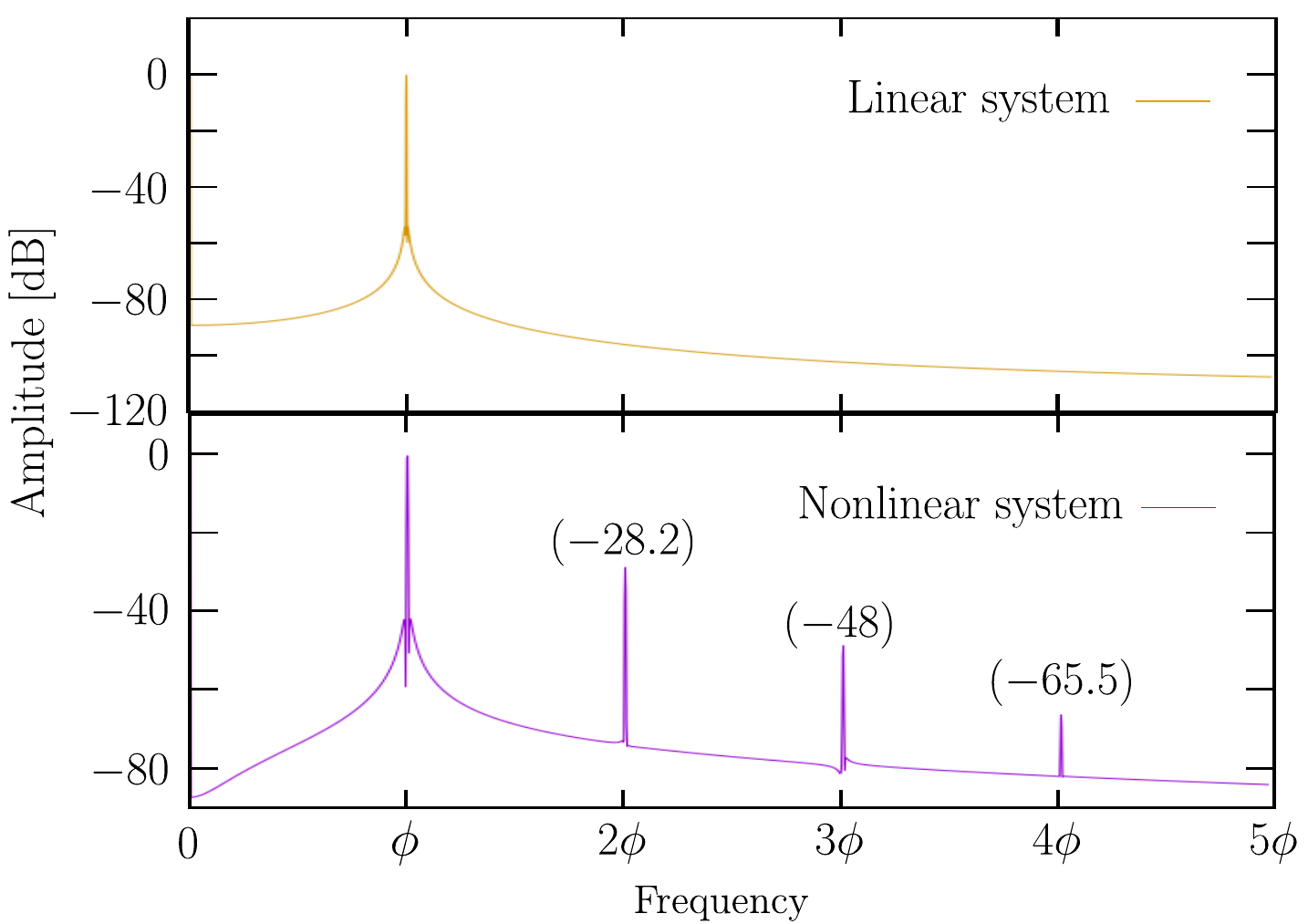}
\caption{\label{fig:2} FFT of the PL intensity simulated as a function of time. PL modulates at a single frequency given by the modulation frequency of the excitation beam for an excitonic system that relaxes with an exponential decay (linear system). The PL in a system with excitons and free carriers is modulated at multiple harmonics of the excitation beam's modulation frequency (nonlinear system). }
\end{figure}

 To illustrate the principle, we compare the frequency components in the PL from a molecular system and a crystal by numerically simulating the response using respective rate equations governing the dynamics of excited species. The details are provided as supplementary information. 
The excited species in molecules are excitons with their recombination given by a simple linear rate equation
  \begin{equation}\label{EQ1}
 	\frac{d n_{\textrm{ex}}}{d t} = g(t) - \frac{ n_{\textrm{ex}}}{ \tau_{\textrm{ex}}},
 \end{equation}   
 where $g(t)$ is the rate of exciton generation by optical excitation and  $\tau_{\textrm{ex}}$ is the lifetime of the excitons.  As the PL signal, $S(t)$, in molecules is proportional to $n_{\textrm{ex}}$, solution to the rate equation for $g(t)\propto I(t)$ gives $S(t) \propto (1+\cos(\phi t))$ for $\phi <<\frac{1}{\tau_{\textrm{ex}}}$, which has a DC component and an AC component exactly at $\phi$. The FFT analysis, shown in Figure \ref{fig:2} (top) for a $\tau_{ex}=5$ ns, confirms a single peak at the modulation frequency. In contrast, optical excitation of a semiconductor at room temperature results in the formation of free carriers and excitons. The nonlinear rate equations for analyzing recombination processes are~\cite{FOKIN1978} 
 \begin{eqnarray}\label{EQ2}
 	\frac{d n}{d t} &=& g(t)-(\gamma+\gamma_{\textrm{ex}}) n^2-\frac{n}{\tau_e}+\alpha n_{\textrm{ex}}+\frac{1}{2} C n^2_{\textrm{ex}} \\\nonumber
 \frac{d n_{\textrm{ex}}}{d t} &=& \gamma_{\textrm{ex}} n^2 -(\alpha+\frac{1}{\tau_{\textrm{ex}}}) n_{\textrm{ex}} - B n n_{\textrm{ex}} - C n^2_{\textrm{ex}}, 
 \end{eqnarray}
 where $n$ is the free-electron population, $\gamma$ is the rate of direct recombination of electrons, $\gamma_{\textrm{ex}}$ is the recombination to form excitons, $\tau_e$ is the electron relaxation time, $\alpha$ is the coefficient of thermal dissociation of excitons, $B$ is the reaction constant of exciton-electron interaction, and $C$ is the exciton-exciton interaction constant. Two of the recombination processes, the direct recombination of electrons and relaxation of excitons, are radiative. Figure \ref{fig:2}(bottom) shows the FFT of the simulated PL for $(\gamma+\gamma_{\textrm{ex}})=1.265\times 10^{-7}$ cm$^3$s$^{-1}$, $\gamma_{\textrm{ex}}=1.0\times 10^{-7}$ cm$^3$s$^{-1}$, $C=1.0\times10^{-7}$ cm$^3$s$^{-1}$, $\tau_e=6.77\times 10^{-8}$ s, $\tau_{\textrm{ex}}=8.7\times 10^{-10}$ s, $\alpha = 6.43\times 10^8$ s$^{-1}$, and $g_{\textrm{peak}}=1.29\times 10^{25}$ cm$^3$ s$^{-1}$. The electron-exciton interaction term is neglected at the low excitation densities used here. Parameter values are chosen via an optimization process to match the experimental results of CdSe as detailed in the Supplementary Information.  The FFT shows additional frequency components at the harmonics of $\phi$ resulting from the nonlinear dynamics. The amplitude at 2$\phi$ is $\sim 4\%$ of the fundamental, implying that the system is highly nonlinear even at modest excitation. 
 
 \begin{figure}
\includegraphics[width=1\linewidth]{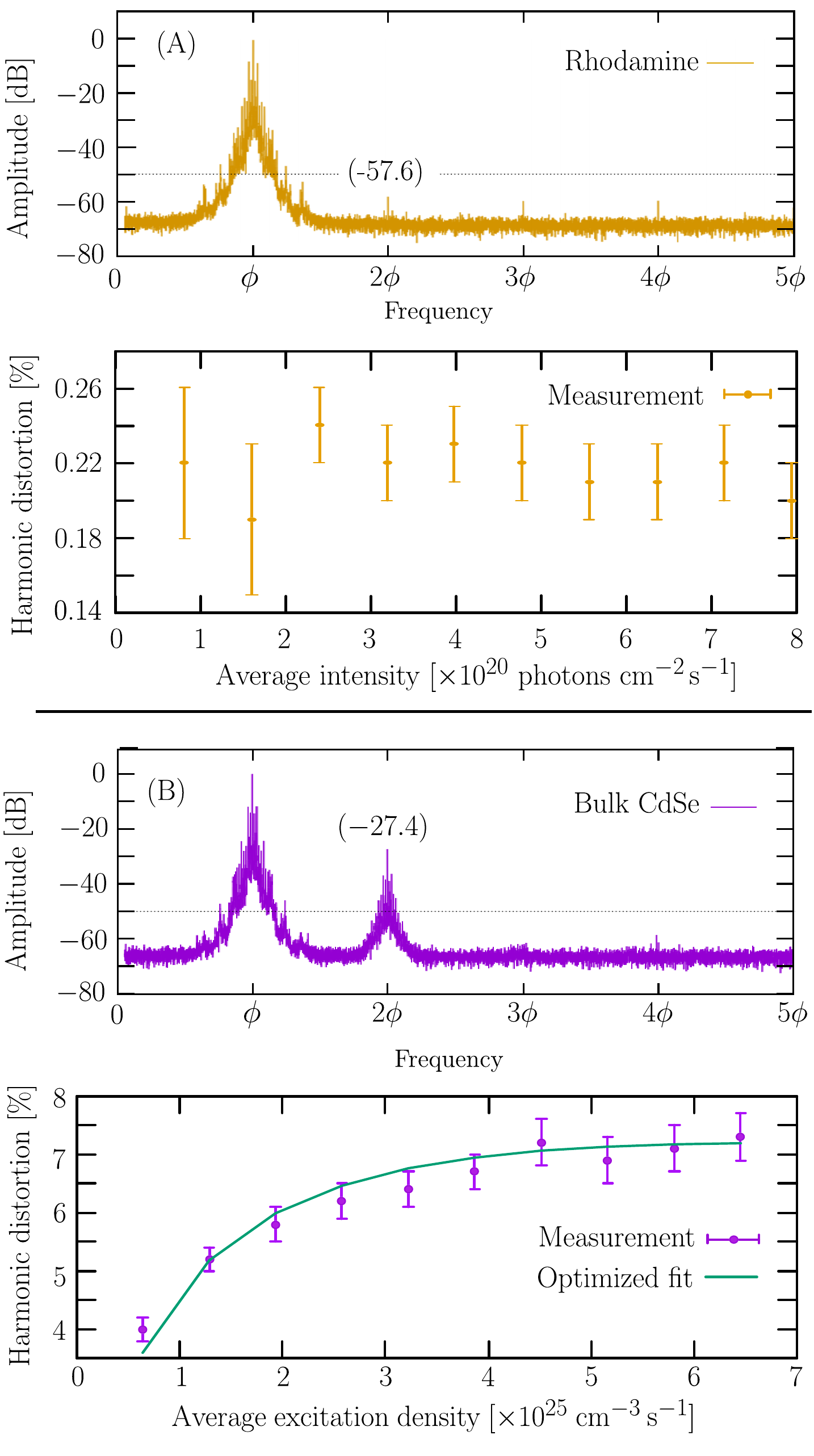}
\caption{\label{fig:3} (A) FFT of the measured PL intensity in RhB (top) and HD as a function of average photon intensity (bottom). (B) Corresponding FFT and HD from the measured PL in CdSe crystal.  }
\end{figure}

Next, we validate our model by comparing the numerical simulations with the experimental data. 
For a linear system, we use PL from 100 $\mu$M Rhodamine B (RhB) in water. Compared to the simulations, the measured signal shown in Figure ~\ref{fig:3} (A, top) exhibits weak modulations at multiple harmonics. Rather than arising from the sample response, these features stem from inherent nonlinearities in the signal detection pathway, as evidenced by their insensitivity to excitation intensity, as shown in Figure~\ref{fig:3} (A, bottom), and by their presence when the APD detects excitation light directly. Consequently, we define an instrumental sensitivity threshold at \(0.3\%\) (\(\sim -50\) dB) of the fundamental peak and disregard any weaker harmonic components. In contrast, the PL from a prototypical emissive semiconductor CdSe shown in Figure~\ref{fig:3} (B, top) displays a dominant response at the modulation frequency alongside a distinct second harmonic with a relative amplitude of \(\sim 4\%\). This prominent second harmonic confirms that the signal originates from a highly nonlinear process, in excellent agreement with our simulations. Although weak higher-order signals are also present, they fall below the detection threshold and are excluded from the analysis. To quantify this nonlinear response, we define the harmonic distortion (HD) as \(\frac{S(2\phi)}{S(\phi)}\times 100\% \). As shown in Figure \ref{fig:3}(B, bottom), the HD at varying average peak generation rates, \(g_{peak}\), increases rapidly at low excitation levels before trending toward saturation. Simulations reveal that the dependence of HD on \(g_{peak}\) is highly sensitive to the parameters in Eq. (\ref{EQ2}); we exploit this sensitivity to optimize and extract the best-fitting parameter values. A comparison between the parameters reported from literature time-resolved PL measurements~\cite{FOKIN1978} and our best-fit values (Table~\ref{tab:table1}) highlights the distinct advantages of characterizing nonlinearities via modulated photoexcitation. Specifically, nonlinear effects on PL transients are typically too subtle to be precisely quantified by standard exponential fitting, as exponential functions lack orthogonality. Consequently, literature values are often restricted to orders of magnitude, and the underlying differential equations are routinely simplified to include only dominant parameters~\cite{FOKIN1978}. Conversely, our results demonstrate that the frequency-domain response is remarkably sensitive to minute nonlinear effects, enabling more refined parameter extraction without requiring any model simplifications.

\begin{table}
\caption{\label{tab:table1} Comparison of the parameters used to fit time-resolved PL in ref.~\cite{FOKIN1978} and the excitation density dependent HD in this work.   }
\begin{ruledtabular}
\begin{tabular}{lcr}
Parameter& Ref.~\cite{FOKIN1978} & Optimized\\
\hline
$g_{\textrm{peak}}$ (cm$^{-3}$ s$^{-1}$) & $10^{28}-10^{29}$ & $10^{24}-10^{26}$\\
$\gamma_{\textrm{ex}}$ (cm$^{3}$ s$^{-1}$) & $10^{-7}$ & $10^{-7}$\\
$\gamma +\gamma_{\textrm{ex}} $(cm$^{3}$ s$^{-1}$) & $10^{-7}-10^{-6}$  & 1.265$\times 10^{-7}$\\
$C$ (cm$^{3}$ s$^{-1}$)& 10$^{-13}$-10$^{-7}$& 10$^{-7}$\\
$\tau_e$ (s) &-&$6.67\times10^{-8}$\\
$\tau_{\textrm{ex}}$ (s) &-& $8.7\times 10^{-10}$\\
$\alpha$ (s$^{-1}$) &-&$6.43\times10^8$\\
\end{tabular}
\end{ruledtabular}
\end{table}

 \begin{figure}
\includegraphics[width=1\linewidth]{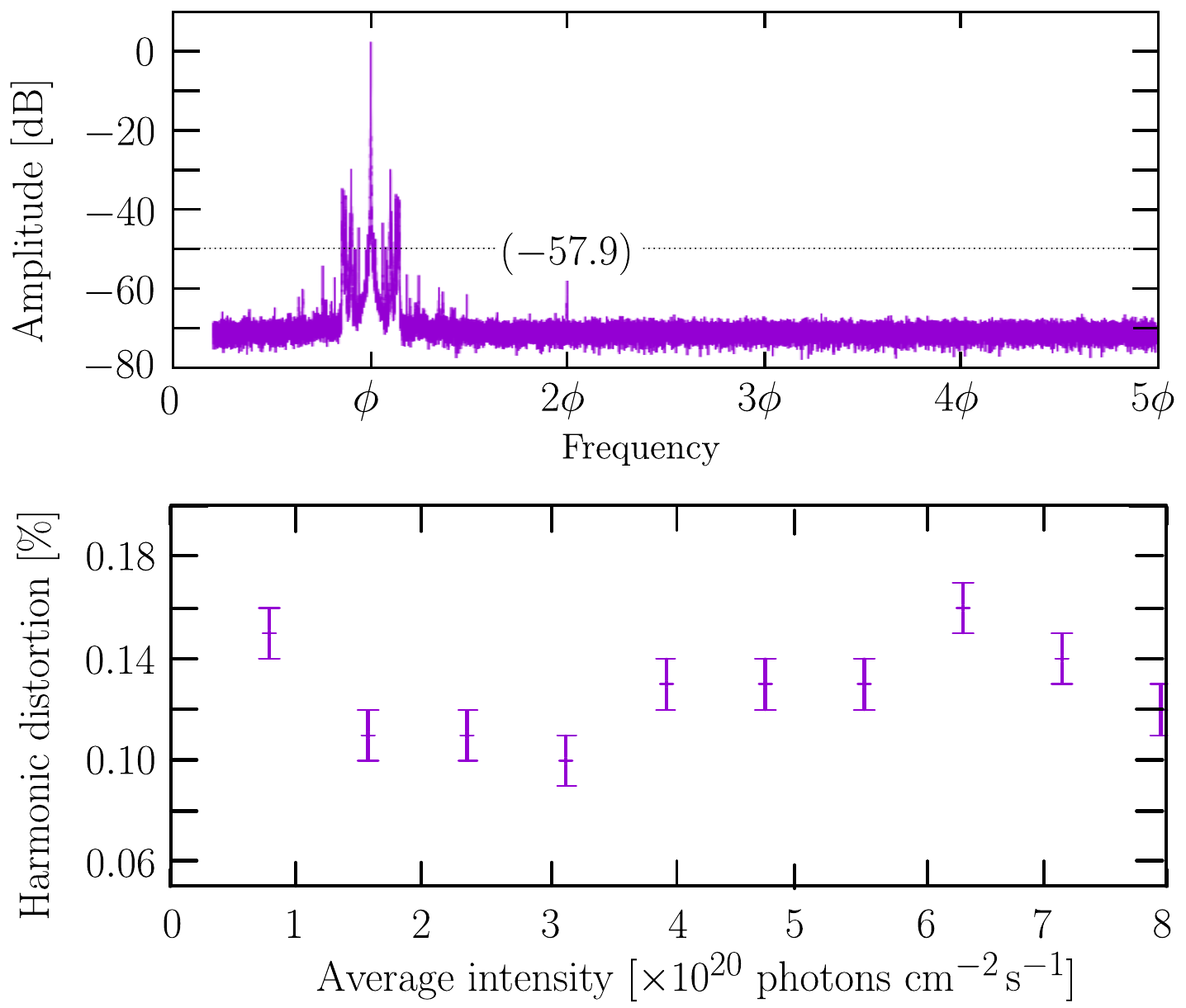}
\caption{\label{fig:4} (A) FFT of the measured PL intensity in CdSe/ZnS QDs (top) and HD as a function of average photon intensity (bottom). }
\end{figure}

As an application of the method, we measure the HD in CdSe/ZnS core-shell quantum dots (QDs)—characterized by a \(94\%\) emission quantum yield at \(620\text{ nm}\)—to investigate how quantum confinement alters relaxation dynamics. The carrier relaxation follows a sequential path: excitation at \(450\text{ nm}\) generates hot electrons (\(n_{\textrm{hot}}\)) and holes above the bandgap, which thermally relax to the bandedge to form excitons. These excitons subsequently return to the ground state via radiative recombination. While exciton recombination operates as a linear process, the preceding thermal relaxation may exhibit nonlinear behavior depending on the experimental conditions~\cite{CAO_2023,LEVINE_2025}, which can be quantitatively probed via HD measurements. FFT of the measured PL signal measured at photon intensity of $\sim 0.8\times 10^{20}$ photons cm$^{-2}$ shown in Figure ~\ref{fig:4} (top) does not exhibit modulated response at $2\phi$ above the instrument noise, indicating that the complete relaxation is a linear process. HD measured at different intensities shown in Figure ~\ref{fig:4} (bottom) are similarly negligible. As the nonlinearity in relaxation of hot electrons results primarily due to multiparticle interactions via three nonlinear processes within a quantum dot, namely the Auger recombination, hot phonon bottleneck, and state filling~\cite{CAO_2023,LEVINE_2025}, we conclude that at intensities of $10^{20}-10^{21}$ photons cm$^{-2}$s$^{-1}$ attained in our CW excitation, none of them are significant. Corresponding photodynamics calculations conform to this (see Supplementary Information).  

In summary, we have demonstrated a clean, highly sensitive frequency-domain method based on phase-modulated CW photoexcitation to characterize nonlinear carrier dynamics in semiconductors. By employing interference between two phase-coherent beams to generate single-tone intensity modulation, this approach bypasses the spectral contamination from excitation overtones that typically limits direct modulation techniques. Analyzing the PL harmonic distortion—specifically the second-harmonic to fundamental ratio—successfully isolates signatures of nonlinear recombination. Applied to bulk CdSe, the extraction of a prominent second harmonic signal (~4\% of the fundamental) enabled the precise numerical optimization of multi-species rate equations, bypassing the non-orthogonality limitations of standard time-resolved exponential fitting. In contrast, application to high-yield CdSe/ZnS core-shell quantum dots revealed a strictly linear relaxation pathway under CW intensities up to \(10^{21}\text{ photons cm}^{-2}\text{s}^{-1}\), confirming that multi-particle nonlinearities like Auger recombination, hot phonon bottleneck, and state filling remain negligible under these operating conditions. The method's key advantages—background-free detection via single-tone intensity modulation, immunity to excitation overtones, and sensitivity to subtle nonlinearities—make it a simple yet powerful diagnostic tool. It is broadly applicable to a wide range of optoelectronic materials and devices, offering a complementary approach to time-resolved spectroscopy for understanding ultrafast nonlinear processes that influence slower, device-relevant responses such as photocurrent and PL.

\begin{acknowledgments}

This work was generously supported by National Key R\&D Program of China (2023YFA1407100), the Quantum Science and Technology-National Science and Technology Major Project (Grant No. 2025ZD0301000), and the Guangdong
Province Science and Technology Major Project: Future
functional materials under extreme conditions (FMUXC, Grant No. 
2021B0301030005). 
\end{acknowledgments}

\section*{Data Availability Statement}
\begin{center}
\renewcommand\arraystretch{1.2}
Data available on request from the authors

\end{center}

\bibliography{aipsamp}

\end{document}